\documentclass[final,5p,times,twocolumn]{elsarticle}
\journal{Physics Letters B}
\usepackage{epsfig}
\usepackage{graphicx}
\usepackage{latexsym}
\usepackage{bm}
\usepackage{float}
\usepackage{color}
\usepackage{fancybox}
\usepackage{hhline}
\usepackage{dcolumn}
\usepackage{textcomp}
\usepackage{epsfig,graphics,graphicx}
\usepackage{amsfonts,amssymb,amsmath}
\usepackage{pifont}
\usepackage{bm}
\usepackage{longtable} 
\usepackage{appendix}
\usepackage{lscape}
\usepackage[mathscr]{euscript}
\usepackage{mathrsfs}
\usepackage{multirow}
\usepackage{rotating}
\usepackage{color} 
\usepackage{placeins}
\usepackage[dvipsnames]{xcolor}

\newcommand{\beq}{\begin{equation}}
\newcommand{\eeq}{\end{equation}}
\newcommand{\bea}{\begin{eqnarray}}
\newcommand{\eea}{\end{eqnarray}}

\newcommand{\ord}[1]{{\cal{O}}\left( #1 \right)}

\usepackage{hyperref}

\def\slashchar#1{\setbox0=\hbox{$#1$}
   \dimen0=\wd0 \setbox1=\hbox{/} \dimen1=\wd1
   \ifdim\dimen0>\dimen1 \rlap{\hbox to \dimen0{\hfil/\hfil}} #1
   \else  \rlap{\hbox to \dimen1{\hfil$#1$\hfil}} / \fi}

\newcommand{\ignore}[1]{}

\begin{document}

\begin{frontmatter}

\title{Heavy baryons in the large $\mathbf {N_c}$ limit}
\author{C. Albertus}
\ead{albertus@ugr.es}
\author{E.  {Ruiz Arriola}}
\ead{earriola@ugr.es}
\address{Departamento de F{\'\i}sica At\'omica, Molecular y Nuclear and
  Instituto Carlos I de F{\'\i}sica Te\'orica y Computacional \\ Universidad
  de Granada, E-18071 Granada, Spain.}
 \author{I. P. Fernando}
\ead{ishara@jlab.org} \author{J. L. Goity}
\ead{goity@jlab.org}    \address{Department of Physics, Hampton University,
  Hampton, VA 23668, USA.}\address{Thomas
  Jefferson National Accelerator Facility, Newport News, VA 23606,
  USA.} 

 \tnotetext[mytitlenote]{Preprint: JLAB-THY-15-2102} 

 \begin{abstract}
It is shown that in the large $N_c$ limit heavy baryon masses can
be estimated quantitatively in a $1/N_c$ expansion using the Hartree
approximation.  The results are compared with
available lattice calculations for different values of the ratio
between the square root of the string tension and the heavy quark mass
$\sqrt{\sigma}/m_Q$.  These estimates implement important $1/N_c$
corrections and assume a string tension independent of $N_c$. Using a
potential adjusted to agree with the one obtained in lattice QCD, a
variational analysis of the ground state  spin averaged baryon mass is performed using Gaussian
Hartree wave functions.  Relativistic corrections through the quark kinetic energy are included. The results provide good estimates for the  first sub-leading in $1/N_c$ corrections. 
\end{abstract}

\end{frontmatter}

\section{Introduction}

QCD in the large $N_c$ limit becomes a non-trivial theory in terms of
an arbitrary and fixed t'Hooft coupling $\lambda= \alpha_s N_c
$~\cite{'tHooft:1973jz}.  In that limit,
baryons~\cite{Witten:1979kh}, unlike mesons, remain as complicated
structures (for a recent review see
e.g. \cite{Lucini:2012gg,Lucini:2013qja} and references therein).
This is the result of the strong coupling of mesons to baryons ${\cal
  O} (\sqrt{N_c)}$, giving baryons a light meson cloud
which contributes to its mass at leading order in $N_c$. In the world
of QCD with only heavy quarks, the meson cloud becomes suppressed in
$\Lambda_{QCD}/m_Q$,  $m_Q$ being the heavy quark mass, and baryonic
states become amenable to a treatment based on non-relativistic
QCD. Thus, heavy baryons are a good laboratory to study the $1/N_c$
expansion. This simpler setting of QCD permits  a straightforward
application of the mean field approach, which will be used in the present work 
and which should provide a good description of baryons in the large
$N_c$ and large quark mass limits.

The quantitative understanding of the $1/N_c$ expansion has become
possible in the light meson sector~\cite{Bali:2013kia}, where meson
masses have been determined in lattice QCD (LQCD) calculations at
different values of $N_c$ and in the quenched approximation, where the
leading ${\cal O}(1/N_c)$ corrections are absent, and moderate
$N_c$ values allow for a safe extrapolation to the large $N_c$  limit. In addition,
estimates based on short distance constraints provide an analytical
understanding of those results~\cite{Ledwig:2014cla}.  More recently,
LQCD calculations of low lying baryon masses for $N_c=3,~5$ and 7
\cite{DeGrand:2012hd,DeGrand:2013nna} have opened the door for a
quantitative test of the $1/N_c$ expansion in baryons as well. Those
pioneering calculations, which are in the quenched approximation, have
quark masses in the light to moderately heavy range.  The present work
is largely motivated by the possibility that such LQCD calculations
could be extended to heavier quark masses, where the  
framework presented here would become  realistically applicable.

In his seminal paper, Witten~\cite{Witten:1979kh} discussed
specifically heavy baryons in the large $N_c$ limit and invoked the
mean field Hartree approximation.  For heavy quarks, it is built from
the simple two-body Hamiltonian, where the interaction is the OGE (one
gluon exchange)~(see \cite{Goity:2004pw} for details) for the short
range part of the interaction. In addition, there are the long range
confining forces, whose effects become suppressed as $m_Q$ grows, and
also short distance radiative corrections must be taken into account
(running of $\alpha_s$) (see \cite{LlanesEstrada:2011kc}). Furthermore, the effects of three-body interactions are of
potential interest; for a recent discussion in the quark model see
Ref. ~\cite{Vijande:2014uma}. They will be discussed briefly in this work. 

At leading order in the $1/N_c$ expansion, the ground state of the
heavy baryon will be described by a wave function which is the direct
product of single-quark wave functions. Since the hyperfine
interactions have spin-flavor non-singlet effects which are
$\ord{1/N_c}$, it is clear that at leading order the spin-flavor state
of the ground state baryons is in the totally symmetric spin-flavor
state, and the baryon has a spin-flavor contracted symmetry
\cite{Dashen:1993as,Dashen:1993ac}, which holds in the limit $N_c\to
\infty$ at fixed quark mass. The effect of removing the center of mass (CM)
motion is sub-leading in $1/N_c$, and can be implemented using
standard   techniques such as the Peierls-Yoccoz
projection (for a review see
e.g. \cite{ring2004nuclear,blaizot1986quantum} and references therein).

The mean field for heavy quarks at large $N_c$ has been studied in
Refs.~\cite{Cohen:2011cw,Adhikari:2013oca} along with possible
implications for baryonic matter. This work builds on that one and
compare to recent lattice calculations for
$N_c=3,~5,~7$~\cite{DeGrand:2012hd,DeGrand:2013nna} after including
some important $1/N_c$ effects such as the CM correction. Brief
discussions of the role of hyperfine splittings as well as the
expected corrections of many-body forces are also given.  A previous
large $N_c$ analysis has been conducted in Ref.~\cite{Cordon:2014sda}.

Note that in order to have low lying baryons with different spins it
is necessary to have more than one flavor of heavy quark. The mass of
the baryon will then have an $\ord{1/N_c}$ hyperfine contribution
(dependent on the spin $S$ of the baryon).  The masses of ground state
baryons take the form of a rotational band,
\begin{eqnarray} 
M_B(S)=N_c
m_0+\frac{C_{HF}}{N_c}(S(S+1)-\frac 34 N_c)+\ord{1/N_c^2} ,
\end{eqnarray} where
$m_0$ and $C_{HF}$ are $\ord{N_c^0}$  and have an expansion in $1/N_c$, and
depend on the quark mass $m_Q$.  The hyperfine independent component
of the baryon mass given by $m_0$ is obtained  by the following combination of baryon masses:
\begin{eqnarray}
m_0&=& \frac{2}{ { N_c^2} ( { N_c}+1)
   ( { N_c}+3)^2} \nonumber \\ &\times& \sum _{S=\frac 12}^{\frac{ { N_c}}{2}}
   \left( 3 + N_c (3 N_c+2) - 8 ( N_c-3) S\right)
    {M_B}\left(S  \right).
\label{eq:m0}
\end{eqnarray}

The baryon masses studied here will be the ones with the hyperfine
effects removed, i.e., $\mathring M_B\equiv N_c m_0$. These will be
later compared with the available LQCD results of
Refs. \cite{DeGrand:2012hd,DeGrand:2013nna,Cordon:2014sda}.

Of course, for any different value of $N_c$  one has a different
theory. Thus, in order to relate them one must assume that some
observables are $N_c$ independent. Actually, on general grounds one
has that:
\begin{eqnarray} \frac{m_0}{\sqrt{
    \sigma}}=F(N_c, \frac{m_Q}{\sqrt{\sigma}}), 
\end{eqnarray} 
where $\sigma$ sets the scale of QCD and can be identified for
instance with the string tension, and $m_Q$ is the heavy quark
mass. $F$ is a universal function $ \ord{N_c^0}$ which admits an
expansion in $1/N_c$, and which for large $m_Q$ can be more
conveniently expressed as $F(N_c, \frac{m_Q}{\sqrt{\sigma}})=
\frac{m_Q}{\sqrt{\sigma}} f(N_c, \frac{m_Q}{\sqrt{\sigma}})$.

The present work goes beyond
Refs.~\cite{Cohen:2011cw,Adhikari:2013oca} by analyzing the main
$1/N_c$ contributions such as the CM effect, and
relativistic corrections, and actually compares to available LQCD
results. For $N_c=3$,  triply heavy baryons have been  studied on the
lattice as a $\Omega_{bbb}$ state~\cite{Meinel:2010pw}, and also 
re-addressed in quark models within several
schemes~\cite{LlanesEstrada:2011kc,Vijande:2014uma,Flynn:2011gf}
which, however, have not addressed larger $N_c$ values.

One important goal on the lattice has been to make the quarks as light
as possible. Actually, quarkonium studies based LQCD proceed always
through the determination of the $\bar Q Q$ potential, and a subsequent
solution of the non-relativistic Schr\"odinger equation~(see
e.g. \cite{Laschka:2012cf}). The present work takes a similar point of
view as a $N_c$-body problem. It should be emphasized   that studying
heavy baryons at varying values of $N_c$ will help with the  understanding of  the
$1/N_c$ expansion in a setting where an analytic approach with small
model dependencies can be applied.

\section{Color singlet states}

The starting point is the Hamiltonian for heavy quarks.  Using
non-relativistic heavy quark field operators $Q(x)$, the Hamiltonian
is given by:
\begin{eqnarray}
H \!\!&=&\!\! \int d^3 x  \left[- \frac{1}{2   m_Q}  Q^\dagger(x)\;\Delta Q (x) 
+m_Q\,Q^\dagger(x)  Q (x) \right]   
\\ ~\!\!\!\!\!\!&+&\!\!\!\!\!\! 
\frac12 \int d^3 x \, d^3 x'\, Q^\dagger (x) \frac{{ \lambda}_a}{2} Q(x) \;
Q^\dagger (x') \frac{{ \lambda}^a}{2} Q(x') V(x-x')~, \nonumber
\end{eqnarray}
where $\lambda^a$ are the $SU(N_c)$ generators in the fundamental
representation, and in perturbation theory $V(r) = \alpha_s /r$ is the
OGE interaction. Here,  only  two-body interactions  are included.
The role of many body interactions is commented below. An equivalent
representation for the case of a heavy baryon is the Hamiltonian
\begin{eqnarray}
H= \sum_{i}\left[ m_Q+ \frac{p_i^2}{2 m_Q}\right]+
\frac14 \sum_{i< j}^{N_c} {  \lambda}_a(i) \otimes {  \lambda}^a(j) V(x_i-x_j)
\label{eq:Hamiltonian}
\end{eqnarray}
The $\lambda \otimes \lambda$ interaction implies exact 
Casimir scaling of the potential energy. Casimir scaling for the $Q\bar Q$
potential holds perturbatively up to two loops (there are three-loop
violations)~\cite{Anzai:2010td} and numerically on the lattice
\cite{Bali:2000un}.

For a colour singlet state the wave function is completely symmetric
in the orbital and spin-flavour quantum numbers, and the baryon
behaves effectively as a bosonic system.  In particular, for ground
state baryons the wave function is the product of a  symmetric spacial wave function and a symmetric spin-flavor wave function and reads as follows:
\begin{eqnarray}
\Psi (x_1 ,\dots , x_N)=  \psi(x_1, \dots, x_N) \chi_{SF},
\end{eqnarray}
where $\chi_{SF}$ is the  spin-flavor wave function. For
excited baryon states, spin-flavor and spatial mixed symmetry states also
occur.  The color matrix elements for arbitrary $N_c$ in
the ground state can be computed as follows. Starting with the
quadratic Casimir operator for the fundamental representation   given
by ($F^a=\lambda^a/2$) 
\begin{eqnarray}
\vec F_q \cdot \vec F_q = \vec F_{\bar q} \cdot \vec F_{\bar q} =
\frac{N_c^2 -1}{2 N_c},
\end{eqnarray}
for a baryon (colour singlet) state one obtains:
\begin{eqnarray}
0 &=& \langle B | (\sum_{i=1}^{N_c} \vec F_i)^2 | B \rangle \nonumber
\\ &=& \langle B | \sum_{i=1}^{N_c} (\vec F_i)^2 | B \rangle + 2
\sum_{i<j} \langle B | \vec F_i \cdot \vec F_j | B \rangle \nonumber
\\ &=& N_c \langle B | (\vec F_q)^2 | B \rangle + N_c(N_c-1) \langle B
| \vec F_q \cdot \vec F_{q'} | B \rangle,
\end{eqnarray}
and likewise for a meson state one obtains:
\begin{eqnarray}
0&=& \langle M | (\vec F_q + \vec F_{\bar q})^2 | M \rangle\nonumber\\
& =& 
2 \langle M | (\vec F_q)^2 | M \rangle +  
2 \langle M | \vec F_q \cdot \vec F_{\bar q} | M \rangle 
\end{eqnarray}
These equations lead to 
\begin{eqnarray}
\langle B | \vec F_q \cdot \vec F_{q'}| B \rangle &=& - \frac12  \left( 1+ \frac1{N_c} \right) \label{eq:ff} \\ 
\langle M | \vec F_q \cdot \vec F_{\bar q} | M \rangle &=& - \frac{N_c^2-1}{2N_c}
\label{eq:ffbar} 
\end{eqnarray}
At very short distances the potential between a heavy quark and
antiquark should be described with perturbative QCD, and approximately
given by an $N_c$-independent expression at leading order (LO) in terms 
of the running  strong coupling $\alpha_s^{N_c}(r)$, 
\begin{eqnarray}
V_{Q \bar Q}^{N_c, {\rm LO}} (r)= -\frac{N_c^2-1}{2N_c} \frac{\alpha_s^{N_c}(r)}{r} = 
\frac1{r} \, \frac{6}{11 \log(r \Lambda_{\overline{{\rm MS}}})} \, . 
\label{eq:Vpert}
\end{eqnarray}
At long distances it is of linear confining
  form and the corresponding string tension $\sigma$ is determined in LQCD.  For $N_c=3$ the $\bar Q
  Q$ potential has been computed in LQCD in the quenched
  approximation~\cite{Necco:2001xg}, and for $N_c>3$ also
  \cite{DeGrand:2012hd,DeGrand:2013nna}. For $N_c=3$, it is well
  described by the bosonic string model~\cite{Luscher:2002qv}, namely:
\begin{eqnarray}
V_{Q \bar Q}^{N_c=3}(r) =   
  -\frac{\pi}{12 r} + \sigma\, r \, . 
\label{eq:Vqqbar}
\end{eqnarray}
The Coulomb term on the RHS is what results from the fluctuations of
the string.  It is remarkable that it provides the bulk of the Coulomb
interaction down to the lattice spacings used in present day
calculations. Using 
$\Lambda_{\overline{{\rm MS}}}/\sqrt{\sigma}=0.503(2)(40)+ 0.33(3)(3)/N_c^2 + 
      {\cal O}(N_c^{-4})$ obtained in \cite{Allton:2008ty} one gets  that at $r
  \sqrt{\sigma} \sim 0.2 $ the $1/r$ term in 
Eqs.~(\ref{eq:Vpert}) and (\ref{eq:Vqqbar}) coincide.  
For the heavy quark mass corresponding to Compton wave lengths much
smaller than present lattice spacings, where the long distance
potential plays a minor role, the Coulomb interaction
will increasingly become the one predicted by perturbative QCD,
Eq.~(\ref{eq:Vpert}).

At arbitrary $N_c$, $V_{Q \bar Q}^{N_c}$ will only receive corrections
$\ord{1/N_c^2}$, as required by the $1/N_c$ expansion in pure
gluodynamics.  Assuming the leading scaling in $N_c$ for $\alpha_s$
and $\sigma$, and Eq~(\ref{eq:ffbar}), the potential becomes: 
\begin{eqnarray} 
V_{Q \bar  Q}^{N_c}(r) &=&
\frac 98 \frac{N_c^2-1}{N_c^2} V_{Q \bar
  Q}^{N_c=3}(r)\nonumber\\
  &=&(1+\ord{1/N_c^2}) V_{Q \bar Q}^{N_c=3}(r).
\label{eq:Vqq0} 
\end{eqnarray} 
This $N_c$ dependence will be loosely named "Casimir scaling". This is
verified by the t'Hooft coupling $\lambda = 4 \pi N_c \alpha_s$ used
in Refs.~\cite{DeGrand:2012hd,DeGrand:2013nna}.  Clearly this follows
only if the above assumption is made,
and with the present calculation
at $N_c>3$ it can be verified,  as discussed below.

\begin{figure}[h]
\begin{center}
\epsfig{figure=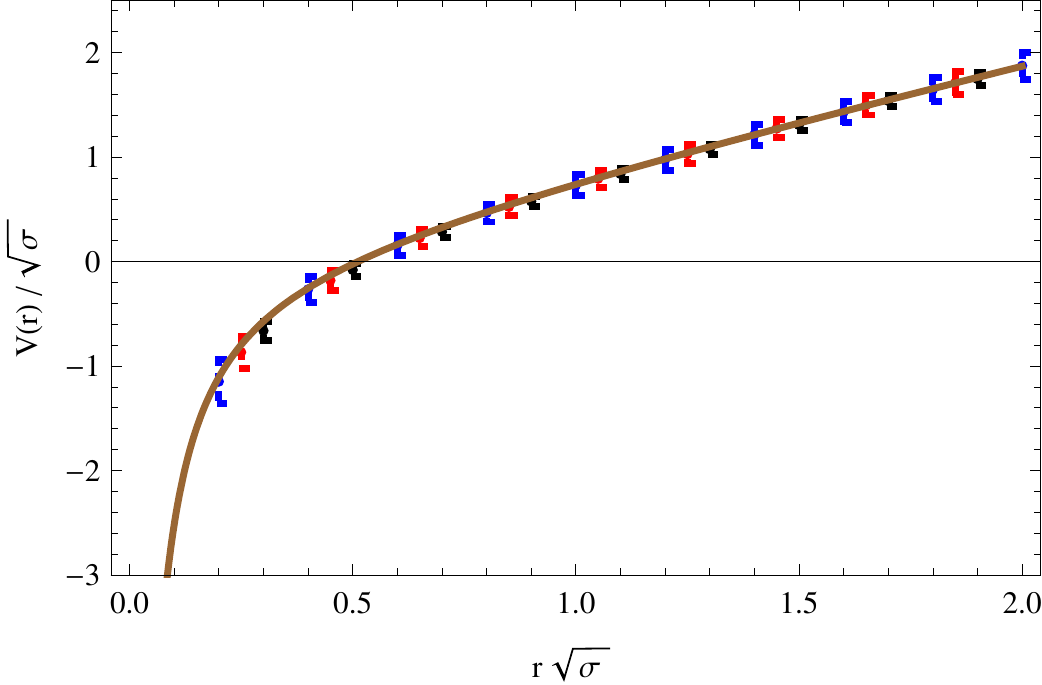,width=8cm}
\end{center}
\caption{Quark-antiquark Potential on the lattice in units of the
  string tension for $N_c=3,5,7$ compared with the bosonic string
  model~\cite{Luscher:2002qv} (full line). The 
  values for different $N_c$:   $3$ (blue),  $ 5$ (red)
  and $ 7$ (black), have been transported to avoid
  cluttering of points.}
\label{fig:potscal357}
\end{figure}

As mentioned earlier, the $1/N_c$ expansion requires definition because it  compares different theories. The most obvious
way to proceed is to require that certain quantities are independent
of $N_c$, e.g., the string tension and quark masses at a given scale.
Since the LQCD results of Ref. \cite{DeGrand:2012hd,DeGrand:2013nna} have the property that the
string tension is approximately independent of $N_c$, i.e.,
$\sigma=\frac 98 \frac{N_c^2-1}{N_c^2}\sigma(3)\sim \text{const}$, this condition
is adopted in what follows. The result from Fig.~\ref{fig:potscal357}
vividly shows the $N_c$ independence  of the $Q\bar Q$ potential within
the current lattice uncertainties and the astonishing agreement with
the bosonic string model~\cite{Luscher:2002qv}. Thus, generalizing the
$N_c$ lattice findings~\cite{Necco:2001xg} the potential for all $N_c$ will be taken to be:
\begin{eqnarray}
V_{Q \bar Q}^{N_c}(r)= V_{Q \bar Q}^{N_c=3}(r)=
-\frac{\pi}{12r} + \sigma\, r 
\label{eq:Vqbarq}
\end{eqnarray}

From Eqs. (10-15) the two-body interaction potential in the baryon
becomes:
\begin{eqnarray}
V_{Q Q}^{N_c}(r) = \frac{V_{Q \bar Q}^{N_c}(r)}{N_c-1} =
\frac{1}{N_c-1}\left(-\frac{\pi}{12r} + \sigma\, r \right)
\label{eq:Vqq}
\end{eqnarray}

\section{Mean field approximation and beyond}

\subsection{Mean field approximation}

The calculation for different values of $N_c=3,~5,~7, \dots$ of the
baryon mass with the Hamiltonian Eq. (\ref{eq:Hamiltonian}) requires
solving separate few body problems with their inherent technical
complications. In the large $N_c$ limit, however, an important
simplification arises as a mean field approach becomes valid.  The
structure of the spacial wave function is of Hartree type
~\cite{Witten:1979kh}, and in the particular case of the ground state
it reads:
\begin{eqnarray}
\psi (x_1 ,\dots , x_N)=  \prod_{i=1}^N \phi(x_i) \, . 
\end{eqnarray}
For a single baryon, the baryon mass    $ \mathring{M}_B=  
\langle \psi | H | \psi \rangle \equiv \langle H \rangle_\psi$ is given by:
\begin{eqnarray}
\mathring{M}_B &=& N_c m_Q + N_c \int d^3 x \frac{1}{2 m_Q} | \nabla \phi (x) |^
2 \nonumber \\ &+& \frac{N_c(N_c-1)} 2 \int d^3 x d^3 x' |\phi(x')|^2
|\phi(x)|^2 V_{QQ}(x-x'). \nonumber \\ 
\label{eq:mfe}
\end{eqnarray} 
The large $N_c$ scaling becomes obvious after the relation,
Eq.~(\ref{eq:Vqq0}) is used.  It is useful to define 
the effective mean field potential $\bar
V(x)$ generated by $N_c-1$ quarks 
\begin{eqnarray}
\bar V(x) &=& (N_c-1) \int d^3 x' V_{QQ}(x-x') |\phi (x')|^2  \nonumber \\
 &=& \int d^3 x' V_{Q \bar Q}(x-x') |\phi (x')|^2 ~,
\label{eq:mean-f}
\end{eqnarray}
where  the Casimir scaling assumption provided by
Eq.~(\ref{eq:Vqq}) has been used. The mean field potential is the self-energy of a
quark within  the hadron which sees the remaining $N_c-1$ quarks
(which are coupled into  the anti-fundamental representation $\bar F$).

The mean field equations are then obtained by minimizing with respect to a
normalized $\phi(x)$ leading to the   eigenvalue problem:
\begin{eqnarray}
-\frac{1}{2 m_Q} \nabla^2 \phi(x) + \bar V(x) \phi (x) = \epsilon \phi (x).
\label{eq:sch}
\end{eqnarray}

\subsection{Numerical and variational solution} 

The mean field equations Eqs.~(\ref{eq:sch}) and (\ref{eq:mean-f}) can
be solved by iterations until self-consistency solution is
obtained. Actually, for the case $\sigma=0$ the system can be written
as a coupled Schr\"odinger-Newton equation, which was already solved
in Ref.~\cite{arriola1999asymptotic}. A Gaussian ansatz of the form 
\begin{eqnarray}
\phi(r) =\left(\frac{2}{\pi b^2}\right)^{\frac34} e^{-r^2/b^2}
\label{eq:gauss}
\end{eqnarray}
yields a good approximation to this solution and allows for a simple
analytical discussion.~\footnote{In the $\sigma=0$ case one has $\mathring M_B - 3
  m_Q= -0.00034 \alpha_s^2 \,m_Q$~\cite{arriola1999asymptotic} vs $\mathring M_B - 3
  m_Q= -0.00031 \alpha_s^2 \,m_Q$ from Eq.~(\ref{eq:gauss}).  For the case
  $\sigma \neq 0 $  more sophisticated ans\"atze were tried embodying
  better short and long distance behaviors, but improvement is at the
  per cent level since the quarks are located in the mid-range
  region.   Discussion of  several possibilities will be  given elsewhere.}

\subsection{CM corrections and mass formula}

One standard and well documented problem of the mean field
approximation in nuclear physics is the violation of Galilean
invariance~\cite{blaizot1986quantum,ring2004nuclear} which is a
symmetry of the starting Hamiltonian, Eq.~(\ref{eq:Hamiltonian}),
namely the invariance under the boost operation with velocity v,
$\Psi(x_1, \dots, x_N) \to e^{i m_Q \text{v} \cdot \sum_i x_i } \Psi(x_1, \dots,
x_N) $, which implies the energy of the
moving system to be given by $E(P)= M+ P^2/2m_QN_c$ where the rest
mass differs from the inertial mass $M \neq N_c m_Q$.

Since the interest here is to include $1/N_c$ corrections in the
calculation, it is important to build a wave function that is an
eigenfunction of the momentum. This is achieved by implementing, e.g.,
the Peierls-Yoccoz projection
method~\cite{blaizot1986quantum,ring2004nuclear}~\footnote{Semiclassical
  collective quantization methods provide an alternative after due
  attention to zero modes is
  paid~\cite{blaizot1986quantum,ring2004nuclear}.}. However, for the
simple Gaussian single particle wave function, Eq.~(\ref{eq:gauss}),
this corresponds just to replace $N_c \to N_c-1$ in the kinetic energy
contribution. Thus,   the projection becomes trivial to deal
with, and one obtains for a moving baryon of momentum $P$:
\begin{equation}
\mathring{M}_B=N_c m_Q+\frac{P^2}{2m_Q {N_c}}+ \frac{3
   ({N_c}-1)}{2{b}^2
   m_Q}
   +\frac{N_c}{b\sqrt{\pi}}\left(- \lambda ^2
   + b^2\,
    \sigma\right), \label{eq:massB}
\end{equation}
where $\lambda^2=\pi/12$.
Minimizing with respect to $b$ ($b_0$) yields the baryon mass at rest.  At
large $m_Q$, $b_0$ and the baryon mass become:
\begin{eqnarray}
b_0&=& \frac{3 \sqrt{\pi } }{\lambda ^2 m_Q}\frac{N_c-1}{N_c}\left(1-9 \pi \left(\frac{N_c-1}{N_c}\right)^2
\frac{\sigma}{\lambda^6 m_Q^2}
\right)\nonumber\\ &+&\ord{1/m_Q^5}\nonumber\\ \mathring M_B&=&N_c m_Q+\frac{P^2}{2
  m_Q {N_c}}\nonumber\\
  &+&({N_c}-1) \left(\frac{3 \sigma }{\lambda ^2
  m_Q}-\left(  \frac{N_c}{N_c-1} \right)^2\frac{\lambda ^4 m_Q}{6 \pi }\right)\nonumber\\
  &+&\ord{1/m_Q^2},
\end{eqnarray}
which shows a delayed onset of the heavy quark regime due to large
numerical factors. Thus, one should expect relativity to play a role
even for moderately heavy quarks.

\subsection{Relativistic corrections}

Of course, a full relativistic treatment implies particle creation as
implied by locality, and Poincar\'e invariant Hamiltonian methods with
a fixed number of particles exhibit well known features (see
e.g. Ref.~\cite{Keister:1991sb} and included references). While this
can be improved,   here only an estimate of the relativistic
corrections is considered  by the standard replacement at the single particle level,
$ m_Q + p_i^2/2m_Q \to \sqrt{p_i^2+m_Q^2} $, which leads remarkably to
an analytical expression for the zero momentum projected variational
energy
\begin{eqnarray}
\mathring M^{\text{rel}}_B&=&
\frac{1}{\sqrt{2\pi} \,b}\,\Big(
\frac{{b}^2 m_Q^2\, {N_c}^{3/2} e^{\frac{{b}^2 m_Q^2 {N_c}}{4
   ({N_c}-1)}} K_1\left(\frac{{b}^2 m_Q^2 {N_c}}{4
   ({N_c}-1)}\right)}{  \sqrt{{N_c}-1}}\nonumber\\
   &+&\sqrt{2}\,N_c \left(-\lambda
   ^2+{b}^2 {\sigma} \right)\Big),
\label{eq:mass-rel}
\end{eqnarray}
which reproduces from the simple non-relativistic CM rule $N_c \to
N_c-1$ in the kinetic energy in the heavy quark limit~\footnote{Note
  that here  one  projects and does not boost the mean field solution. In the
  relativistic case the rest and inertial masses ought to coincide due
  to Poincar\'e invariance. The necessary identity between boosting
  and projecting onto linear momentum only holds for exact
  solutions~\cite{Betz:1983dy}.  At the mean field level the identity
  is guaranteed at the mean field solution~\cite{Pobylitsa:1992bk}.}.
The scheme as in the mean field case of minimizing with respect to the
oscillator parameter $b$ yields the final baryon mass at any
$N_c$~\footnote{Note that the direct extrapolation of
  Eq.~(\ref{eq:mass-rel}) to light quarks $m_q \to 0$ leads to the rest
  mass $ \mathring M_B /( N_c\sqrt{\sigma})= 1.81 - 0.50 /N_c - 0.19/N_c^2 +
  \dots $, which is the crude estimate  for the multiplet center
  in the quenched approximation.}. This case will be used in order to
compare with the LQCD results in
Ref.~\cite{DeGrand:2012hd,DeGrand:2013nna}, where the largest quark
masses used are still not in the heavy regime.

\subsection{Ground state correlations} 

As expected Eqs.~(\ref{eq:sch}) and (\ref{eq:mean-f}) are $N_c$
independent and correspond to the leading order approximation. These
equations have corrections corresponding to different physical
effects. Within the Gaussian ansatz for the single particle states
Eq.~(\ref{eq:gauss}) a Harmonic oscillator shell model interpretation
applies since the baryon is in a $(1s)^{N_c}$ state. In this picture,
ground state correlation correspond to virtual excitations to higher
shell states $(n_1 l_1) \dots (n_{N_c} l_{N_c})$.

In order to quantify the accuracy of the Hartree approximation within
the large $N_c$ framework, one  evaluates  the variance of the Hamiltonian
defined by $\Delta H_{\psi}^2=\langle H^2\rangle-\langle H\rangle^2$
where $\langle O \rangle \equiv \langle \psi | O | \psi \rangle
$. When solving the equation approximately, as it is done here using a
variational wave function, it turns out that $\Delta H_{\rm
  var}/\langle H \rangle =\ord{1/\sqrt{N_c}}$ typical of statistical
fluctuations.  Straightforward calculation,  explicitly
using the mean field equation  Eq. (\ref{eq:sch}), shows that~\footnote{Here the notation corresponds to
\begin{eqnarray} 
\langle V_{QQ'}\rangle \!\!\!\!\!\! \!\!\!&\equiv& \!\!\!\!\!\! \!\!\!\int d^3x \,d^3y\, V_{QQ}(x-y) |\phi(x)|^2 |\phi(y)|^2 \nonumber \\ \langle
V_{QQ'}V_{Q'Q''}\rangle \!\!\!\!\!\! \!\!\!&\equiv& \!\!\!\!\!\! \!\!\!\int d^3x\, d^3y\, d^3z\, V_{QQ}(x-y)V_{QQ}(y-z)
|\phi(x)|^2 |\phi(y)|^2 |\phi(z)|^2 \nonumber 
\end{eqnarray}} 
\begin{eqnarray} 
\Delta H_{\psi}^2= \frac{N_c(N_c-1)}{2} \left[  \langle  V_{QQ'}\rangle^2 + \langle
V_{QQ'}^2 \rangle-2\langle V_{QQ'} V_{Q'Q''} \rangle \right] \, .
\end{eqnarray}
Only when the self-consistent Hartree mean field equation is exactly
satisfied and due to the Casimir scaling assumption,
Eq.~(\ref{eq:Vqq}), one has $\Delta H_{\rm Hartree}=\ord{N_c^0}$,
which  means $\Delta H_{\psi}/\langle H \rangle =\ord{ {1/N_c}}$
for the correction relative to the baryon mass.

\subsection{Multiquark interactions}

In general, there are multiquark interactions which contribute to the
baryon mass at the nominal leading $\ord{N_c}$. For heavy quarks one
expects that in the baryon only n-body interactions with $n\leq N_c$
are of any significance. For $N_c=3$ there is a long history of
studying the 3-quark interactions, where there are two competing
alternatives to confining forces of quarks in baryons, the $\Delta$
(pairwise triangle shape) and the $Y$ (junction shape) inspired by
string models~\cite{Bali:2000gf}.

Three body interactions have been addressed perturbatively
\cite{Brambilla:2009cd} for arbitrary $N_c$. In the present case, the
non-perturbative effect of 3-body interactions can be visualized with
one example. Consider a 3-body potential of the form:
\begin{eqnarray}
  V_3(x_1,x_2,x_3)=\sum_{i=1}^3 \text{v}_3(x_i-X) \; d_{abc}\;
  \lambda^a\otimes\lambda^b\otimes\lambda^c, 
\end{eqnarray} 
where $X$ is the CM position of the three quarks. The expectation
value of $V_3$ in the baryon ground state at rest can be evaluated
explicitly choosing $
\text{v}_3(r)=\frac{1}{N_c^2}\left(-\frac{\lambda_3}{r}+\sigma_3
\,r\right)$ where $\lambda_3$ and $\sigma_3$ are $\ord{N_c^0}$, one
obtains for the Gaussian wave function:
\begin{eqnarray} \langle
V_3\rangle=2\sqrt{\frac{3}{\pi}} \left(N_c-\frac{5}{
  N_c}+\frac{4}{N_c^3}\right) \left( -\frac{\lambda_3}{b}+\sigma_3 \,b
\right) ~,
\end{eqnarray} 
where the color matrix element for
the baryon was used, 
\begin{eqnarray} 
\langle d_{abc}\;
\lambda^a\otimes\lambda^b\otimes\lambda^c\rangle=4
\frac{(N_c-3)!}{N_c!}(N_c^3-5 N_c+\frac{4}{N_c}) .
\end{eqnarray} 
Note that the expectation value of the 2-body interaction
Eq.~(\ref{eq:massB}) and the one of the 3-body interaction studied here
have the same form except that their $N_c$ scalings differ by terms
which are of relative order $1/N_c^2$. Therefore, the 3-body forces
cannot be distinguished from the 2-body ones unless those higher order
terms in the expansion are taken into account. This is in a sense
direct consequence of the mean field approximation, which naturally
"hides" the n-body nature of the interactions.  Other
n-body forces are in principle possible for a large $N_c$ baryon,  whose color structure is
given by $1/N_c^{n-1}\;d_{a_1\cdots a_n} \lambda^{a_1}\otimes\cdots\otimes
\lambda^{a_n}$, where $d_{a_1\cdots a_n}$ is the rank  $n$  invariant
symmetric tensor of $SU(N_c)$. A simple calculation shows that they
contribute to the baryon mass with an overall factor $N_c/n!$, which
implies that even for very large $N_c$,  $n$-body forces with $n>5$
become very suppressed.

\begin{figure}[t]
\begin{center}
\includegraphics[width=8cm]{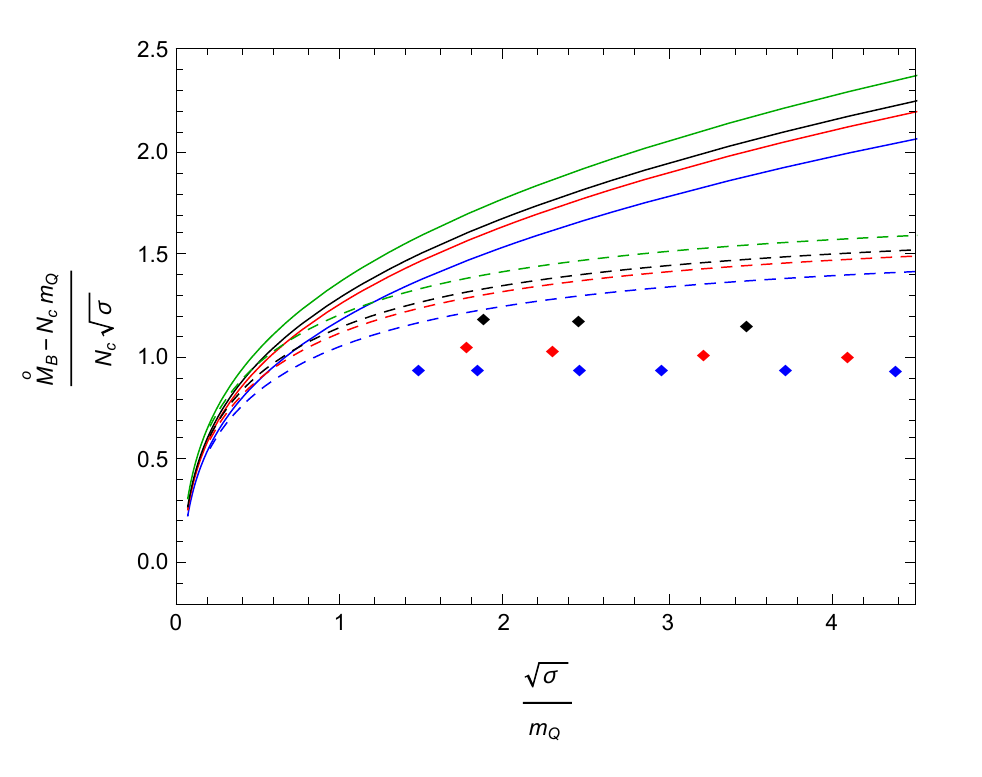}
\end{center}
\caption{Baryon mass as a function of the string tension for Gaussian
  wave function.  Depicted are the results for non-relativistic (full
  curves) and relativistic (dashed) calculations, and the lattice QCD
  results for $m_0$ defined by Eq.~(\ref{eq:m0})
  (diamonds)~\cite{Cordon:2014sda}. The color coding is that of Fig. 1, and in green the limit $N_c\to \infty$. The
  string tension corresponding to the lattice QCD results was obtained
  as explained in the text.}
\label{fig:baryon-mass}
\end{figure}

\begin{figure}[t]
\begin{center}
\includegraphics[width=8cm]{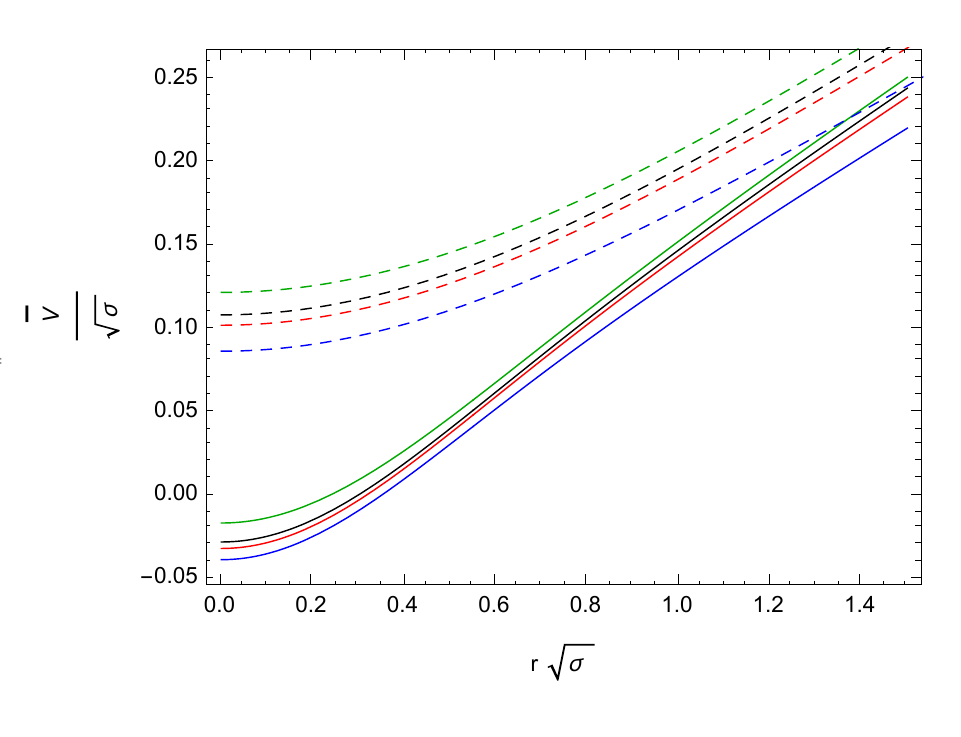}
\end{center}
\caption{Effective potential for a heavy quark in a heavy baryon with $m_Q=2\sqrt{\sigma}$ (dashed) and  $10\sqrt{\sigma}$ (full).  Same color coding as in Fig.~\ref{fig:baryon-mass}.}
\label{fig:eff-pot}
\end{figure}

\subsection{Hyperfine effects}

The simple OGE potential contains hyperfine components ${\cal O}
(m_Q^{-2})$, which have implications on meson spectra (see
e.g. Ref.~\cite{Segovia:2011tb}), as they contribute at $\ord{N_c^0}$
in mesons, but contribute to hyperfine splitting in baryons only at
$\ord{1/N_c}$. They can be easily evaluated as perturbations using the
wave function obtained here.  A quick calculation generalizing the
$N_c=3$ result \cite{Isgur:1979be} to arbitrary $N_c$ gives for the
hyperfine mass shifts: \beq \delta M_B^{HF}(S)=\frac{8}{3\sqrt{\pi}}
\frac{\alpha^{N_c}_s(m_Q)}{m_Q^2 b^3}(S(S+1)-\frac 34 N_c).  \eeq They
play no role for the spin-weighted average baryon mass
Eq.~(\ref{eq:m0}).

\section{Towards relating to LQCD results}

Following the motivation of this work, the aim here is to compare the
mean field description including relativistic and CM corrections with
results from LQCD.  At present, the only available LQCD results for
ground state baryon masses at several $N_c$ values are those of
Refs.~\cite{DeGrand:2012hd,DeGrand:2013nna} (slightly updated in
Ref. \cite{Cordon:2014sda}), where quenched calculations have been
undertaken at several values of the quark mass and at $N_c=3,~5,$ and
7. While the purpose there was to pursue the light quark limit, here
the opposite situation is emphasized where simplifications are
expected and the quenched approximation is better fulfilled.

As discussed earlier, the explicit $N_c$ dependence is inferred from
taking $\sigma$ to be $N_c$ independent.  The lattice results
displayed in
Refs.~\cite{DeGrand:2012hd,DeGrand:2013nna,Cordon:2014sda} are given
in lattice units, with $a$ the lattice spacing.  Using the form of the
quark-quark potential the Sommer parameter $r_1$ is determined by the
standard definition
\begin{eqnarray}
-r_1^2 V_{Q \bar Q}'(r_1)= -1 ~,
\end{eqnarray}
yielding in the present  case 
\begin{eqnarray}
  r_1^2 \sigma = 1 - \frac{\pi}{12}~.
\end{eqnarray}
This value, namely $r_1 \sqrt{\sigma}=0.859$, is roughly valid for the
LQCD calculations with $N_c=3,~5 ~\text{and} ~7$, where the respective
results from Table I of Ref. \cite{DeGrand:2012hd} are 0.856(5),
0.850(4) and 0.845(2). Using the values of $r_1/a$ in the same Table
one obtains respectively $\sqrt{\sigma}
\,a=0.219(2),~0.225(2),~\text{and}~0.216(1)$. For the level of
precision of the present comparison it is therefore sufficient to take
$\sqrt{\sigma} \,a=0.22$ for all $N_c$.  While the main goal
of~\cite{DeGrand:2012hd,DeGrand:2013nna} was to pursue the lowest
quark mass limit, some moderately high quark masses were included.
These are now used to compare with the results of this work.

The numerical results are presented in Fig.~\ref{fig:baryon-mass}. As expected, the relativistic limit sets in at about $\sqrt{\sigma} \sim
m_Q$. The lattice data of Ref.~\cite{DeGrand:2012hd,DeGrand:2013nna} stop at twice
larger values, so it would be highly interesting to extend the lattice
calculations to the non-relativistic regime, where the theory can be
more easily handled.

The mean field approximation is visualized through the mean field
potential $\bar V(r)$ created by the $N_c-1$ quarks, see
Eq.~(\ref{eq:mean-f}). In the present  case, for zero momentum states and the
Gaussian profile, Eq.~(\ref{eq:gauss}) one obtains:
\begin{eqnarray}
\bar V (r) &=&  
\frac{b_0\,\sigma}{\sqrt{2\pi}}\, e^{-\frac{2 r^2}{b_0\,^2}}  \nonumber \\ 
&-&  \left(
\lambda ^2-\sigma \left(\frac{b_0^2}{4}+ r^2\right)\right)
\,\frac 1r \; \text{erf}\left(\frac{\sqrt{2} r}{b_0\,} \right),
 \end{eqnarray}
which is shown for illustration, in Fig.~\ref{fig:eff-pot} for
different values of $N_c$ and $m_Q$. Improvements to this behavior
correct for long distance behavior and will be discussed in a
forthcoming publication.

\section{Conclusions}

In the present work, a scheme is put forward where the large $N_c$
expansion of baryon masses in the lattice can be described in terms of
the mean field approximation as originally advocated by Witten and
$1/N_c$ corrections thereof. The quark-quark potential is assumed to
follow Casimir scaling at arbitrary $N_c$ and hence proportional to
the quark-antiquark potential, which to good accuracy as per current
LQCD calculations is $N_c$-independent.  This provides a universal
$N_c$ independent scheme where the ratio of the baryon mass to $N_c
\sqrt{\sigma}$ can be numerically evaluated.

It was shown that the corrections to the mean field energy are
generically ${\cal O} (\sqrt{N_c})$, but become ${\cal O} (N_c^0)$,
when the mean field energy takes its minimum value.  This accuracy is
the result of the density of quarks in the baryon growing as
proportional with $N_c$.  Among the estimated corrections are the
leading in $N_c$ relativistic ${\cal O}(m_Q^{-3})$ and subleading
${\cal O} (N_c^0)$ CM corrections. Hyperfine splittings are removed by
suitably averaging over spin states.  When compared with available
LQCD calculations, the present results account within $20\%$ for the
dimensionless ratio $( \mathring M_B- N_c m_Q)/( N_c \sqrt{\sigma})$
which is of natural size. This is encouraging, as it suggests to push
the LQCD calculations to heavier quark masses and also refine the
calculations in the present work.

One of the obvious benefits of the present investigation is the
possibility of going beyond the ground state and extend these ideas to
the excited baryon spectrum, where lattice calculations are admittedly
more involved and less accurate.  LQCD calculations of excited baryons
for $N_c>3$ may still be an unreachable goal. However, it is likely
that this will be achieved first with heavy quarks, and in that case
the approach followed here can be easily used to predict the excited
states.  Finally, other heavy baryon properties, such as form factors, are easily derived with the wave functions obtained here.
\section{Acknowledgments}
Useful discussions with Thomas DeGrand and correspondence with Marco
Panero are greatly appreciated. This work was supported in part by DOE
Contract No. DE-AC05-06OR23177 under which JSA operates the Thomas
Jefferson National Accelerator Facility (J.~L.~G.), by the National
Science Foundation through grant PHY-1307413 (I.~P.~F. and J.~L.~G.)
and the Spanish Mineco (grant FIS2014-59386-P) and Junta de Andaluc\'ia
(grant FQM225) (C.~A.~T. and E.~R.~A.).   C.~A.~T. acknowledges a
contract from the CPAN.

\bibliographystyle{elsarticle-num}
\bibliography{heavy-largeNc}

\end{document}